\def\aas{{A\&AS}}
\def\aj{{AJ}}

\def\apj{{ApJ}}

\def\mnras{{MNRAS}}

\documentclass[cup5b]{caps}
\usepackage{graphicx}
\usepackage{amssymb}
\usepackage{ociwsymp3e}  %Use this for contributed papers.

\HeadText{E. R. Carrasco, C. Mendes de Oliveira \& L. Infante}  %Enter author name; if three authors, list all three;
\begin{document}

\pagenumbering{arabic}

%Author names should be in captital letters
\author[]{E. R. CARRASCO$^{1}$, C. MENDES DE OLIVEIRA$^{2}$ and L. INFANTE$^{3}$
\\
(1) Gemini Observatory/AURA, Casilla 603, La Serena, Chile\\
(2) IAG-USP,Caixa Postal 3386, 01060-970, S\~ao Paulo, Brazil\\
(3) Dept. de Astronom\'{\i}a y Astrof\'{\i}sica, PUC, Casilla 104, Santiago 22, Chile}

%Example for multiple authors:
%
%\author[]{D. C. BACKER$^{1}$, A. H. JAFFE$^{2}$, and A. N. LOMMEN$^{3}$
%\\
%(1) University of California, Berkeley, CA, USA\\
%(2) Imperial College, London, England\\
%(3) Sterrenkundig Instituut ``Anton Pannekoek'', Amsterdam, The Netherlands}
%

\chapter{The low surface brightness dwarf members of nearby groups of galaxies}

\begin{abstract}

We present the photometric and spectroscopic results of a survey done in a wide-field area 
of nearby groups and poor clusters. The main goals are identifying and characterizing 
the properties of low-surface-brightness dwarf galaxies and determining the galaxy luminosity 
function for M$_V > -13$. We surveyed areas (typically 0.1--1.0 degree$^2$) of the groups 
NGC 6868 (Telescopium), NGC 5846, HCG 42, Hydra I, Dorado and the poor clusters IC 4765 in V and I.
In this work we present the results and the analysis of the photometry and spectroscopy done
in four groups: HCG 42, NGC 6868, NGC 5846 and IC4765. In these groups, we detected hundreds of 
new low surface brightness dwarf galaxies. The analysis of their spatial distributions, colors,
galaxy luminosity function and other propierties are presented. 

\end{abstract}

\section{Introduction}

Many of the galaxies in the Universe are concentrated in low-density environments 
like groups. Except for the Local Group and other nearby structures (e.g. M81, Sculptor, 
Leo I), little has been done to characterize the galaxy population in these environments. 
Group of galaxies contain few bright galaxies with known redshifts and for magnitudes fainter 
than  $M_V=-16$, the galaxy population is poorly know, basically because the spectroscopic 
redshift for low-luminosity (dwarf) galaxies are difficult to obtain. This had a direct impact 
in the determination of the luminosity function at fainter limits, where the dwarf galaxies 
dominate the galaxy counts.

Dwarf galaxies (dE, dSph, dIrr) are the most common type of galaxies in the Local 
Universe. They are thought to be the single systems with the largest dark-matter content, 
with M/L ratios as high as that of groups and poor clusters (Carignan \& Freeman, 1988). Hence 
their spatial distribution and mass spectrum may give us important insight into the spatial 
scales over which mass is distributed in the Universe. However, because of their low luminosity 
and low surface  brightness our understanding of these objects is rather limited. 
Dwarf galaxies in nearby clusters and groups have been studied in detail with photographic material 
(e.g. Impey et al. 1988, Ferguson \& Sandge 1991). With the advent of large-format CCDs, the
searches for low-surface brightness dwarf members have become possible in wide-field CCD images 
of nearby groups and clusters (e.g. Carrasco et al. 2001, Carrasco 2001, Trentham \& Tully 2002).

The main goal of our project is to identify the population of low-surface-brightness dwarf (LSBD) 
galaxies in nearby groups to M$_V =-10$ mag and determine the luminosity function in each of
the groups. The selected sample is formed by groups with different environments, poor groups like
Dorado and NGC 6868 and rich groups like Hydra I and IC 4765 with galaxy populations of different 
sizes and morphological mixtures. The program was started with the observation of the Dorado group 
(Carrasco et al. 2001a), at $cs\sim1200$ km/s, and  continued with the observation of other
seven nearby groups with  $1000<cz<4500$ km/s. The aim of this paper is to present the results of 
the study of the dwarf galaxy population in four of the groups: HCG 42, IC 4765, NGC 5846 
and NGC 6868.

\section{Data acquisition and reductions}

The groups were observed in February and March 1998, with the 1.3m Warsaw telescope (Las Campanas  
Observatory, Chile), using the standard Johnson V and Cousins I filters. All data were taken
under photometric conditions except for one of the fields in NGC6868, with a seeing ranging from 
1'' to 1.3''. For each group, except for NGC6868, two or three images just outside the groups 
were taken as a control fields. The targets were observed with a small  overlapping  region between 
the fields in order to check the photometry and estimate the  photometric errors. The images were
bias/overscan-subtracted, trimmed and flat fielded  using standard procedures. The zero point 
calibrations were obtained using standard stars from Landolt (1992).

The spectroscopic observations were done in July 1999 and March 2000, with the 2.5m
Du Pont telescope at Las Campanas Observatory, using the the Wide Field CCD camera (WFCCD) in 
multi-object mode. The spectra were observed with a low resolution red grism (wavelength coverage
$\lambda \sim 3800-7600$\AA, dispersion $\sim 4$\AA/pix, resolution $\sim 8$\AA). All galaxies
with $V\le21$ mag from the photometric catalogs, including the LSBD galaxies detected in the groups
(see below), were selected for spectroscopic observations. Tipically, 30 to 40 spectra per mask were 
observed.

Table\ref{table1} shows the main parameters of observed groups. The average velocity were calculated using
the bi-weigthed estimators (Beers, Flynn \& Gebhardt 1990) and using all galaxy members with known 
redshift from this work and from the literature. The two last column in the table are the distance 
($H_{0}=75$ km/s/Mpc) and the radius where the density is 200 times the critical density ($R_{200}$).

\begin{table}
\caption{Main parameters of the observed groups}
\begin{center}
\begin{tabular}{lccrrcccc}
\hline \hline
 Group  &  $\alpha(2000)$ & $\delta(2000)$ & $N_{tot}$ & $N_{grp}$ & $V_{avg}$   & $\sigma_{r}$      & D             & $R_{200}$     \\  
        &  [hh mm ss]     &  [dd mm ss]    &           &           &  [km/s]      &   [km/s]         &[h$^{-1}$ Mpc] & [h$^{-1}$ Mpc]\\ \hline
HCG 42  & 10 00 11.4      & --19 37 02     & 204       & 36        & 3828$\pm$51 & 211$^{+30}_{-43}$ & 54.5          & 0.67 \\
NGC5846 & 15 06 28.5      &$+$01 35 10     & 134       & 32        & 1950$\pm$75 & 416$^{+43}_{-63}$ & 24.1          & 1.18 \\
IC 4765 & 18 47 32.9      & --63 18 46     & 167       & 94        & 4488$\pm$65 & 620$^{+40}_{-50}$ & 60.9          & 1.66 \\
NGC 6868& 20 09 43.0      & --48 17 08     &  57       & 19        & 2844$\pm$57 & 238$^{+29}_{-49}$ & 32.7          & 0.50 \\
\hline \hline
\end{tabular}
\end{center}
\label{table1}
\end{table}

\section{Results}

\subsection{Photometry}

The detection, photometry and classification of the objects were performed using the Source Extractor 
(SExtractor) software program (Bertin \& Arnouts 1996). Before running SExtractor, we removed all bright
galaxies, saturated stars and diffuse light from the fields. After sky subtraction, all objects with a 
threshold $\ge$ 1.1$\sigma$ in V ($25.8$ mag/arcsec$^2$) and $\ge$ 1$\sigma$ in I ($\sim 24.4$ mag/arcsec$^2$)
above the sky  level and with a minimum area of 10 pix$^{2}$  ($\sim1.8$ arcsec$^{2}$) were found and extracted
(the V images were used as a templates for the detection of the objects).

The selection of the LSBD galaxies was done using the the parameters given by the exponential profile fit
(central  surface brightness, the scale length and the limiting diameter). The fit to a pure exponential law was 
used to obtain the extrapolated central surface brightness and the scale length of the dwarf galaxy candidates in 
the sample. These two parameters are used as a primary cut of the data. LSBD dwarf galaxy candidates with $\mu_{0}\ge22$ 
V mag/arcsec$^{2}$ and scale length  $h>1.5$'' were selected.  The surface brightness cut is similar to that used by 
Impey et al. (1988) to search for LSBD galaxies in Virgo. The second cut was done  selecting galaxies with a limiting 
angular diameter larger than 1.2 h$^{-1}$ kpc at the given isophotal level. 
These limiting diameters are similar to those used to select the LSBD galaxies in the Dorado group. We used also an 
optimization method (smoothed filter, see the details in Carrasco et al. 2001) to search for galaxies with large profiles 
and very low surface brightness galaxies.  

\begin{figure}
\centering
\includegraphics[width=5.5cm,angle=-90]{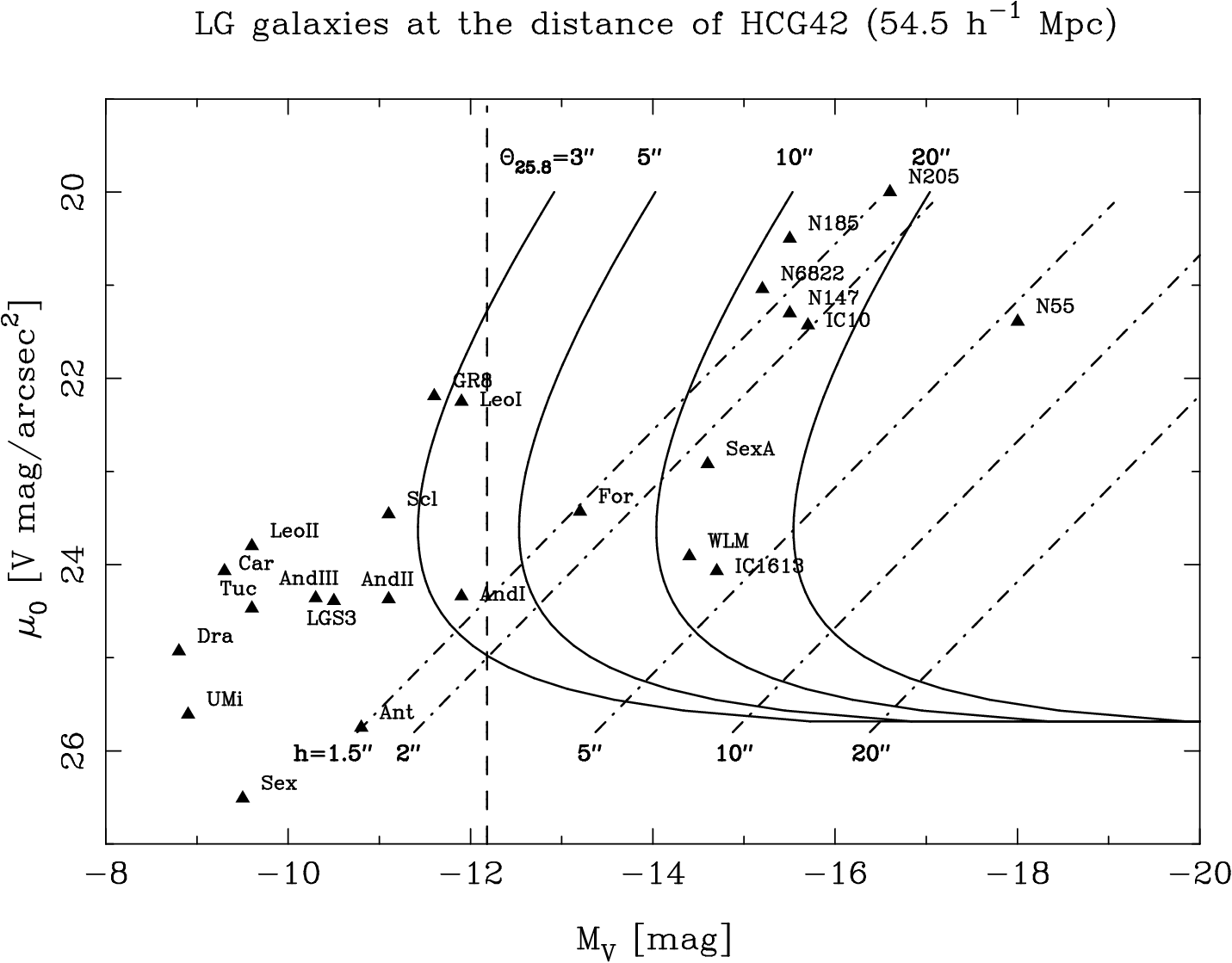}%
\includegraphics[width=5.5cm,angle=-90]{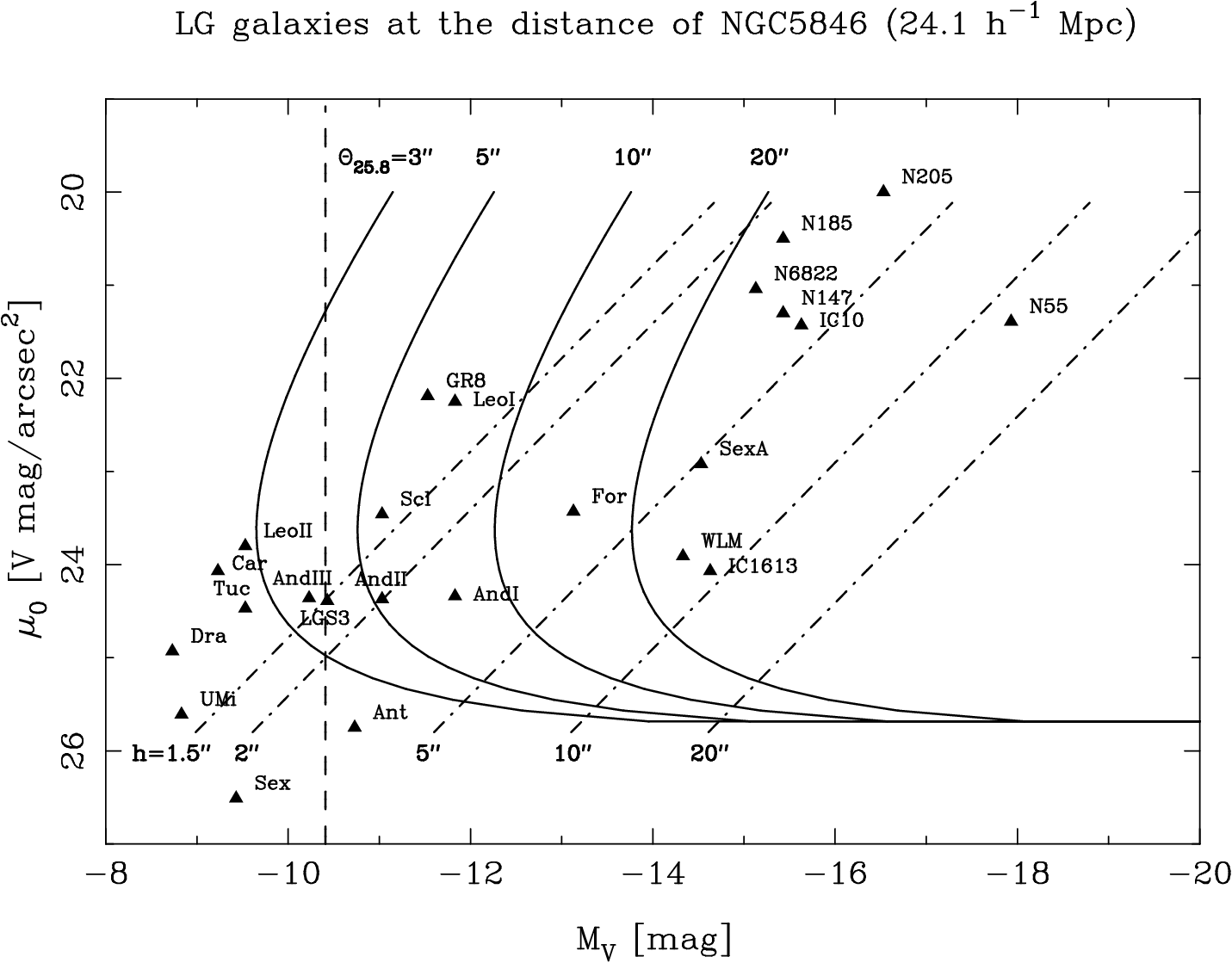}
\caption{$\mu_{0}-M_{V}$ plane for the LSBD galaxies detected in HCG 42 (left) and NGC 5846 (right) groups.}
\label{fig1}
\end{figure}

Ninety new LSBD galaxies were detected, for a cut in diameter of 1.2 h$^{-1}$ kpc. Figure \ref{fig1} shows the $\mu_{0}-M_{V}$ plane 
for the LSBD galaxies detected in HCG 42 and NGC 5846. For comparison, we also plotted  Local Group dwarf galaxies 
at the distance of these two groups (not shown here) where we can see that galaxies like Carina, Draco 
and Umi would not be detected in our survey but Fornax and Sculptor–-like galaxies were detected.  

Monte Carlo simulations were performed on the science frames in order to have the same characteristics as a real image.
The computed completeness fraction of the sample for galaxies with magnitudes and sizes typical of LSBD found 
at the redshift of HCG42 and NGC5846 is $\sim 80$\% for galaxies with $22.5< \mu_{0} <24.5$ V mag/arcsec$^2$ 
and $V<20$ mag. Using the smoothed filter (optimization, see above), the completeness fraction increases from 
$\sim 60$\% to $\sim 80$\% for $V<20$ mag. The differences in magnitude and in the cetral surface brightnesses 
are lower than 0.3 mag. However, the scale length output values are, on average, larger than the input ones by 
$\sim 20$\% .

\subsection{Spectroscopy}

Radial velocity measurements for $\sim 400$ galaxies were obtained. Of these, only 78 are members of the groups. 
We also identified new structures behind the groups. In particular, HCG 42 shows two well– defined background structures 
with $cz=54541\pm202$ km/s, $\sigma_r=853^{+103}_{-159}$ km/s (HCG42-Back1, $N_{mem}=26$) and $cz=72981\pm101$ km/s, 
$\sigma_r=311^{+42}_{-43}$ km/s (HCG42-Back2, $N_{mem}=16$). In addition, the distribution of the galaxies in  HCG42 
suggests the presence of two voids at $cz\sim30000$ km/s and $cz\sim44000$ km/s. The poor cluster IC 4765 also shows a well 
defined background with a $cz=10975\pm125$ km/s, $\sigma_r=461^{+66}_{-115}$ km/s (IC4765-Back1,$N_{mem}=16$).

\subsection{Galaxy Luminosity Function}

\begin{figure}
\centering
\includegraphics[width=5.5cm,angle=-90]{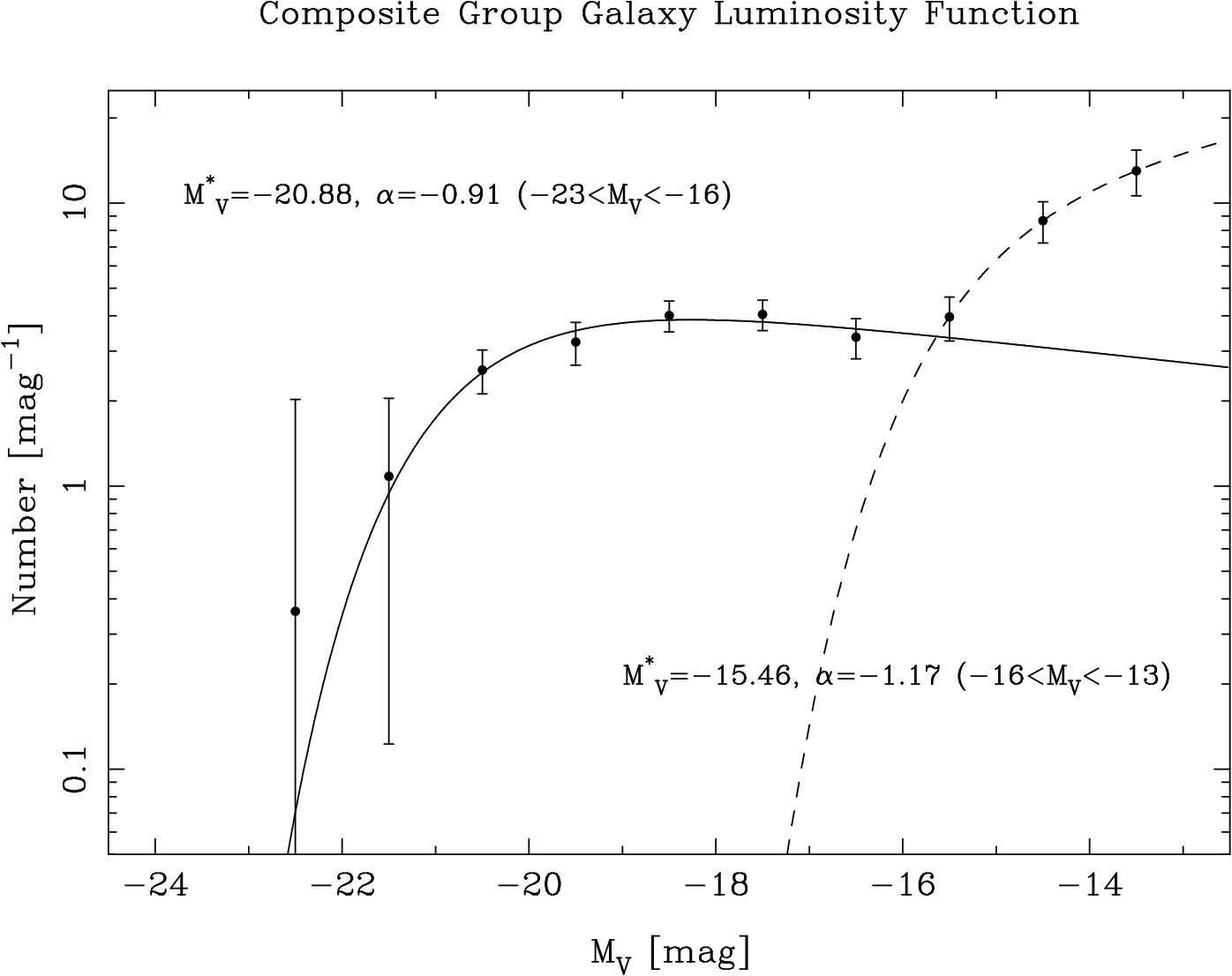}%
\includegraphics[width=5.5cm,angle=-90]{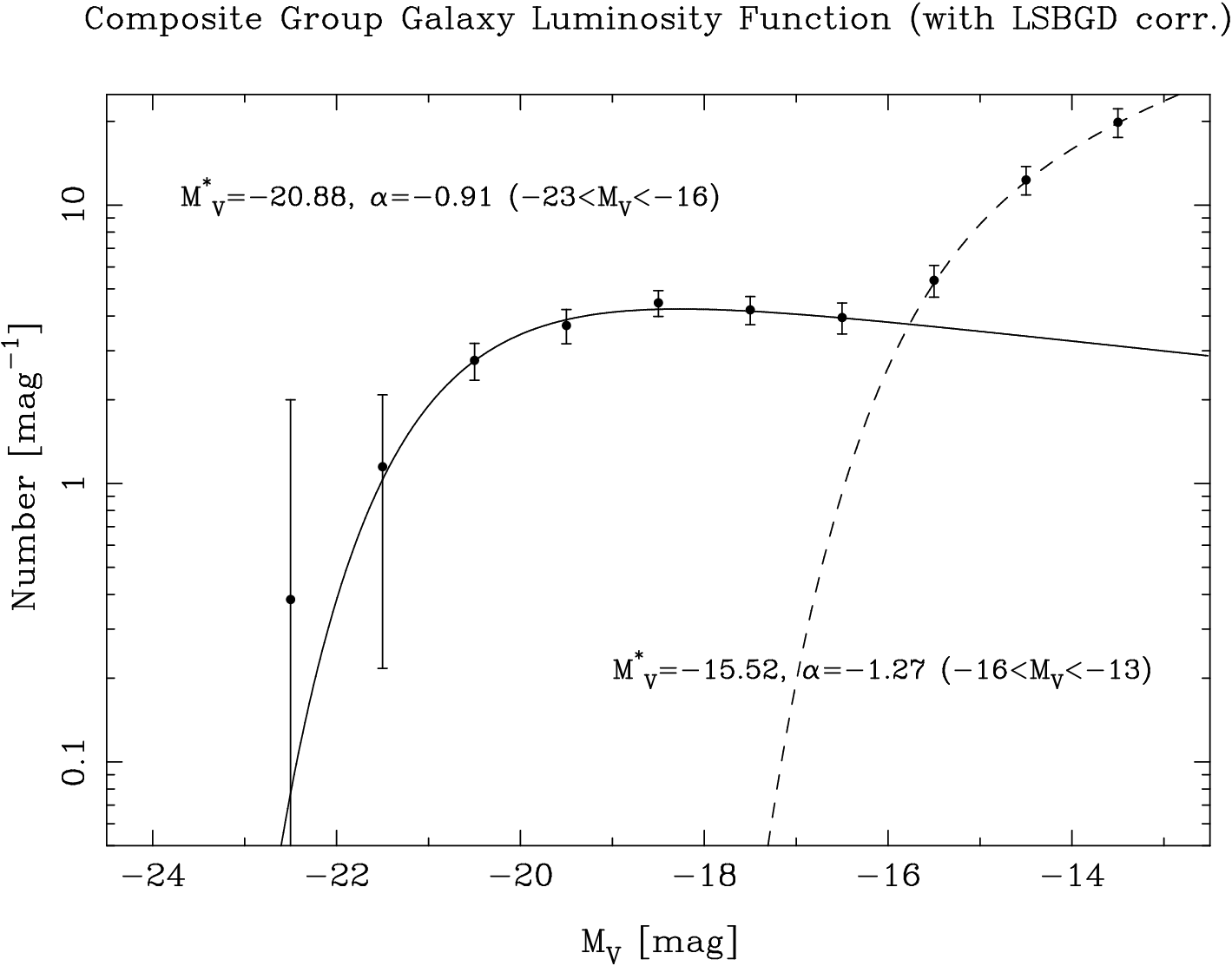}
\caption{Composite Group Luminosity Function (CGLF) for group sample. Left panel: 
CGLF without the correction for incompleteness in the detection of the low-surface 
brightness dwarf galaxies. Right panel: CGLF with the correction in the detection of 
the low-surface brightness dwarf galaxies. }
\label{fig2}
\end{figure}

We determined the using group galaxies with and without corrections for incompleteness in the detection of the low-surface brightness 
dwarf galaxies. Figure \ref{fig2} shows the Composite Galaxy Luminosity Function (CGLF) for the groups without the completeness correction (left) 
and with the completeness correction (right plot) in the detection of LSDBs (the completeness corrections were calculated using Monte 
Carlo simulations, see above). The results show that a single Schecter function does not provide a good fit to the data, i.e. the 
luminosity function shows a bi-modal distribution, with a similar faint-end slope of  $\alpha\sim-1.2$ The results are in agreement 
with the results found in the Local Group and other groups and clusters (e.g. Pritchet \& van den Bergh 1999; Zabludoff \& 
Mulchaey 2000; Trentham \& Tully 2002).

\section{Summary}

Ninety new LSBD galaxies were detected in the group sample. Following Carrasco et. al (2001), extensive simulations 
were done to determine the completeness of our sample and characterize  the success of detection. We find that galaxies like 
Sculptor and Fornax are easily detected at the distance of the groups. Spectroscopic redshifts of $\sim400$ galaxies in the  
field of the HCG42, NGC5846, IC4765 and NGC6868 groups were obtained. The results show that only 78 galaxies are members of the 
groups. Galaxies up to 100000 km/s were identified in our spectroscopic survey. Several new structures were identified behind 
the groups. In particular, in the region of the HCG42 we found two new clusters and a filamentary distribution of the galaxies 
up to 75000 km/s. Two voids are clearly presented in the distribution. We find that the projected radial distribution for the 
dwarf red and blue galaxies is the same. However, the distribution in the velocity space shows differences, suggesting that the 
two population occupy different orbits (blue low surface brightness galaxies show a larger peculiar motion than the red population).
Finally, The composite luminosity functions, determined using group galaxies with and without corrections for incompleteness in 
the detection of the low-surface brightness dwarf galaxies show that a single Schecter function does not provide a good fit to the 
data, given a bi-modal distribution, with a similar faint-end slope. This faint-end slope found in this and other 
works is difficult to reconcile with the slope predicted by the Cold Dark Matter theory ($\alpha\sim-1.2$).

\section{Acknowledgments}
The authors are grateful to the polish astronomers and to the Las Campanas Observatory
staff for generous allocation of telescope time at 1.3 meter Warsaw telescope and
for the support. ERC acknowledge the support for this work provided by FAPESP project Nrs. 
96/04246-7 at the time of his PhD. ERC also acknowledge the Gemini Observatory/AURA for
the financial support.

\end{document}